\documentclass[aps,prl,showpacs,twocolumn,superscriptaddress]{revtex4}

\usepackage{epsfig}
\usepackage{graphicx}
\usepackage{amsmath}
\usepackage{amssymb}
\usepackage{amsfonts}
\usepackage{dcolumn}
\usepackage{color}

\begin{document}
\title{Coherent Atomic Soliton Molecules}

\author{Chenyun Yin}
\affiliation{Department of Applied Mathematics and Theoretical Physics,
University of Cambridge, Cambridge, CB3 0WA, United Kingdom.}

\author{Natalia G. Berloff}
\affiliation{Department of Applied Mathematics and Theoretical Physics,
University of Cambridge, Cambridge, CB3 0WA, United Kingdom.}

\author{V\'{\i}ctor M. P\'erez-Garc\'{\i}a}
\affiliation{Departamento de
Matem\'aticas, E. T. S. I. Industriales, Universidad de Castilla-La
Mancha 13071 Ciudad Real, Spain.}

\author{Valeryi A. Brazhnyi}
\affiliation{Centro de F\'{\i}sica do Porto, Faculdade de Ci\^encias, Universidade do
Porto, 
R.~Campo Alegre 687, Porto 4169-007, Portugal}

\author{Humberto Michinel}
\affiliation{\'Area de \'Optica, Facultade de Ciencias, Universidade de Vigo, As Lagoas s/n, Ourense, E-32004 Spain.}

\begin{abstract}
We discuss the dynamics of interacting atomic bright solitons and dark bubbles in bulk immiscible Bose-Einstein condensates.  
Coherent matter-wave clusters can be constructed using dark-bright 
pairs with appropriate phases. In two dimensions we describe novel types of matter-wave molecules 
without a scalar counterpart that can be seen as bound states of vector objects.
\end{abstract}


\maketitle

\emph{Introduction.-} One of the most remarkable achievements in quantum physics in the last decade was 
the experimental realization of coherent atomic matter waves with Bose-Einstein condensates  
(BECs) of ultra-cold alkaline atoms \cite{reviews}. Because of the inter-atomic interactions, BECs support different types of quantum nonlinear coherent excitations: dark \cite{dark}, 
bright  \cite{bright}  and gap atomic solitons \cite{gap}, vortices, vortex rings and related structures \cite{vortices},   
shock waves \cite{shockwaves}, different types of vector solitons \cite{vector}, and others \cite{Faraday}. 
Many other nonlinear phenomena have been theoretically studied in BECs \cite{Kevrekidis}.

A challenging open problem both in BEC experiments and in nonlinear science at large is the construction of complex molecule-like 
coherent matter-wave structures. A possible strategy, that we explore in this letter, is to 
use atomic solitons and bubbles as bricks to construct stable matter wave aggregates which display 
phase-dependent properties. Related problems have been considered in nonlinear optics in relation with the propagation 
of optical beams in  media with saturable nonlinearities \cite{multisolitons}, it being very difficult to construct even metastable
 long-living soliton clusters.  Nonlocal interactions allow for the creation of soliton clusters but local interactions such as those present 
 in ordinary BECs  pose great difficulties \cite{nonlocal}

In the field of matter-waves, trains with a finite number of quasi-one dimensional bright solitons
 have been observed to be robust and stable \cite{bright} due to the presence of the trap \cite{Gerdjikov}. 
However,  the idea does not work for higher dimensions due to the blow-up phenomenon \cite{bright}. 
With repulsive nonlinearities,  dark solitons always repel each other and 
cannot form  bound states \cite{Gerd1}.

In this letter we will show how multicomponent homonuclear Bose-Einstein condensates in the immiscible regime 
allow for the construction of robust solitonic molecules. These matter-wave clusters 
display phase-dependent properties due to their coherent nature.

\emph{Physical system.-}  We will consider two-component homonuclear BECs with atoms in two different hyperfine states 
$|1\rangle$ and  $|2\rangle$ ~\cite{spinor}. We will work in the immiscible regime and consider droplets of atoms in 
component $|2\rangle$ to be phase separated from a component $|1\rangle$ assumed to have a much 
larger number of particles.

\emph{1-D atomic soliton molecules.-} BECs tightly confined along two transverse directions are 
quasi-one dimensional and ruled in the mean field limit by the equations \cite{reduction,BA}

\begin{equation} \label{eq:6}
i \frac{\partial u_j}{\partial t} = -\frac{1}{2} \frac{\partial^2 u_j}{\partial x^2} 
+ \left(\sum_{k=1,2}g_{jk}|u_k|^2-\mu+\Delta_j\right) u_j,
\end{equation}
for $j=1,2$. Without loss of generality we will take the adimensional chemical potential $\mu =1$, and 
$\Delta_1=0$. Immiscibility implies that $g_{12}^2 > g_{11} g_{22}$. The normalization for $u_2$ is given by
 $\int_{\mathbb{R}} |u_2|^2 = 2 (a_{22}/a_0)N_2$, $a_{22}$ and $N_2$ being the s-wave 
scattering length and number of atoms in $|2\rangle$. Finally $a_0 =\sqrt{\hbar/m\omega_{\perp}}$ 
is the  length-scale in which  the adimensional spatial units are measured.  Since we are interested on
the bulk dynamics as in ring-shaped condensates no longitudinal traps will be considered.

Let us first study the case of equal interaction coefficients, $g_{ij} = 1, i,j=1,2$, that is very close 
to the realistic situation e.g. in Rb \cite{spinor} or Na \cite{Na}. The phenomena to be described in this paper are
not dependent on this specific choice of parameters and indeed, later we will consider the effect of 
tunning interactions \cite{Co}. In the former case, Eqs. (\ref{eq:6}) have explicit dark-bright soliton 
solutions
\begin{subequations}\label{DBS}
\begin{eqnarray}
u_1 & = & i\sqrt{\mu}\sin \alpha + \sqrt{\mu}\cos \alpha \tanh \left(\kappa\left[x-q(t)\right]\right),  \\
u_2 & = & \sqrt{	\frac{N_B \kappa}{2}} e^{i\phi} e^{i\Omega t} e^{i\kappa x \tan \alpha}\text{sech}\left(\kappa\left[x-q(t)\right]\right).
\end{eqnarray}
\end{subequations}
where $\kappa = \sqrt{\mu \cos^2 \alpha+(N_B/4)^2}-N_B/4$ is the inverse of the wave packet length, 
$\Omega = \kappa^2(1-\tan^2\alpha)-\Delta_2$, and the soliton center position is 
$q(t) = X_0 + t \kappa \sin \alpha$ \cite{BA}.
\begin{figure}
\epsfig{file=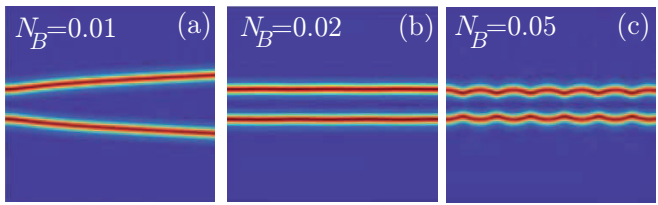,width=\columnwidth}
\caption{Interaction of two vector dark-bright solitons (1,2) for different amplitudes of the bright 
component $N_{B,1}=N_{B,2}$ (a)  0.01 (b) 0.02 (c) 0.05. Other parameter values are given in the text.  
We plot pseudocolor plots of $|u_2(x,t)|^2$ (``bright" component) on the range $x \in [-20,20], t\in [0,2000]$.\label{prima}}
\end{figure}

The repulsive nature of the interaction between dark solitons prevents the formation of bound states of dark solitons.
Concerning bright solitons, their mutual forces in the absence of external effects depend on the phase differences, $\Delta \phi=|\phi_1-\phi_2|$,  going from repulsive for $\Delta \phi = 0$ to attractive for $\Delta \phi = \pi$. 
There is a critical intermediate 
regime for $\Delta \phi$ in which unstable bound states can be constructed. 

\begin{figure}
\epsfig{file=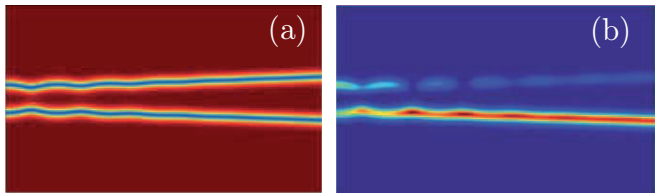,width=\columnwidth}
\caption{Interaction of two vector dark-bright solitons  for $N_B=0.05$ and $\Delta \phi = 3$. 
Other parameters as in Fig.~\ref{prima}. Shown are pseudocolor plots of (a) $|u_1(x,t)|^2$ 
and (b) $|u_2(x,t)|^2$ on the range $x \in [-20,20], t\in [0,2000]$.\label{phasensitive}}
\end{figure}

Thus, a vector object including a dark soliton in one component and a bright soliton in another 
component could lead to a stable bound state with a second vector soliton of the same type when 
the coherent  interactions between the dark and the ``droplet-like" bright component have 
opposite directions leading to what can be considered as a coherent molecule.

We have studied numerically the interaction between two (hereafter 
denoted 1 and 2) initially static ($\alpha_1=\alpha_2=0$) vector solitons given by Eqs.~(\ref{DBS}). 
Other parameters are $\mu=1$, $\Delta_2 = 0$ and the initial positions 
are $X_{0,1} =  3$, $X_{0,2} = -3$. In order to have attractive interactions between the bright components 
we choose $\Delta \phi = \pi$. Our results are summarized in Fig.~\ref{prima}.
\begin{figure}
\epsfig{file=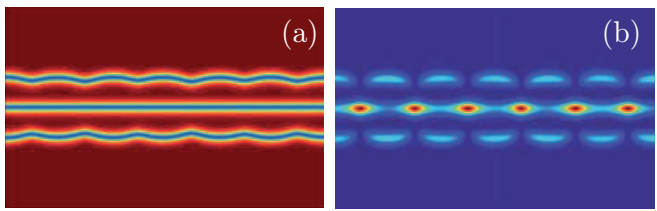,width=\columnwidth}
\caption{Interaction of three vector dark-bright solitons for $N_B=0.05$, $X_{0,1} =-6$, $X_{0,2}=0$, $X_{0,3}=6$. Other 
parameters as in Fig.~\ref{prima}. Shown are pseudocolor plots of (a) $|u_1(x,t)|^2$ and (b) $|u_2(x,t)|^2$ 
on the range $x \in [-20,20], t\in [0,2000]$.\label{4bound}}
\end{figure}

Small droplets of atoms of the second species are not able to stop the outgoing motion 
induced by the repulsive interaction between the black solitons in $|1\rangle$. However, for larger numbers
of atoms in $|2\rangle$  we obtain a bound state of the two vector solitons, with a threshold number of atoms of about $N_B = 0.18$. This is a remarkable result leading to a bound \emph{pair of dark-bright solitons in a system where 
all the interactions are repulsive} due to the effective attraction provided by the phase between the ``droplets" 
of atoms in component $|2\rangle$.
\begin{figure}
\epsfig{file=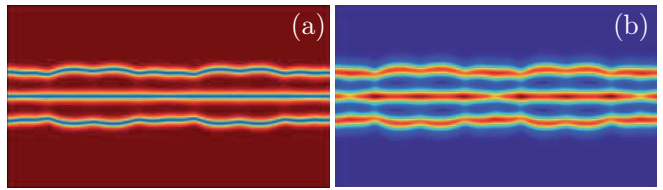,width=\columnwidth}
\caption{Interaction of three vector dark-bright solitons for $N_B=1$, $X_{0,1} =-8$, $X_{0,2}=0$, $X_{0,3}=8$ with $\phi_1=0$, $\phi_2=2.28$ , $\phi_3=0$. Other parameters as in Fig. \ref{prima}. Shown are pseudocolor plots of 
(a) $|u_1(x,t)|$ and (b) $|u_2(x,t)|$ on the range $x \in [-30,30], t\in [0,1000]$.\label{3bound}}
\end{figure}

The formation of these bound states is sensitive to the specific phase difference between the bright droplets in 
$|2\rangle$ as it is seen in Fig.~\ref{phasensitive}, where a change of the phase difference from $\pi$ to 3 suffices to destabilize the system.

However, the addition of more components leads to a very interesting feature: although the atoms in $|2\rangle$ feature 
a less trivial dynamics, they are still able to sustain the multisoliton bound state provided the beating period is faster 
than the typical evolution times of the dark component. An example is shown in Fig. \ref{4bound}.

Curiously, multisoliton states are much more robust to perturbations in the phase differences 
(see  Fig. \ref{3bound}). We have also constructed atomic soliton molecules with more than 
three atoms, e.g. five soliton bound states, etc.

\emph{Two-dimensional molecules.-} When passing to higher dimensions the phenomenology changes 
essentially due to the fact that the more natural building blocks are vortices for the first component hosting a 
``droplet" of the second component. In the scalar case, moving two-dimensional bound states 
were found in Ref. \cite{jr} and correspond to vortex pairs of opposite circulation and rarefaction pulses. A similar 
phenomenology arises in the two-component case in the miscible regime \cite{ngb}. As the velocity increases, the
distance between vortices of opposite circulation decreases to zero. The solutions at even higher velocity are localized 
density perturbations without zeros of the wavefunction. In two dimensions the sequence of solutions terminates 
with solutions approaching zero energy and momentum as the velocity $U$ approaches the speed of sound. 

Due to the structure of the vortex velocity field it is not possible 
to use the interactions of the bright part to construct bound states of equally charged vortices. Thus, unless a simple extension of
the previous ideas is not possible we will see in what follows that a novel 
phenomenology arises very different from the classical one described in Ref. \cite{jr} that can be linked to the presence 
of bound states of solitary wave structures.

\begin{figure}
 \epsfig{figure=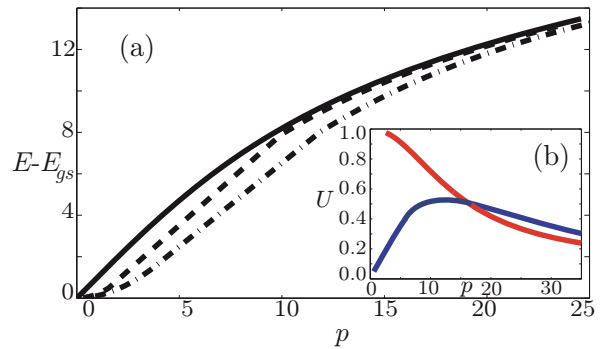, width=0.9\columnwidth}
\caption{(Color online)  (a) Energy ($E-E_{gs}$)-- impulse ($p=p_1+p_2$) dispersion curve of the solitary
  wave solutions of Eqs. (\ref{ueq}) for $\alpha=1.2$ and various
 choices of $N_2$. The upper solid line corresponds to $N_2=0$, i.e. the 
 JR dispersion curve, shown here for comparison. To ease comparison we subtract the ground state energy $E_{gs}$ in the plot. 
  Dashed black line  corresponds to $N_2 = 8$ and the dash-dot line corresponds to $N_2= 20$.
 (b) Velocity as a function of momentum for $N_2= 20$, the red curve corresponding to the JR case.
 The solid line corresponds to the JR case.}
\label{fig_disp}
\end{figure}

In what follows we will systematically construct solitary waves moving with speed $U$ along the $x-$direction 
in two-dimensional two-component condensates in the phase-separation regime as solutions of the coupled GP system \eqref{eq:6}, 
where $\partial/\partial t \rightarrow -U\partial/\partial x$, and $\partial^2/\partial x^2 \rightarrow \nabla^2$
\begin{subequations}
\label{ueq}
\begin{eqnarray}
i U \frac{\partial \psi_1}{\partial x} &=& \frac{1}{2}\nabla^2 \psi_1+
\left(1-|\psi_1|^2 - \alpha |\psi_2|^2\right)\psi_1,
\\
i U \frac{\partial \psi_2}{\partial x} &=&\frac{1}{2} \nabla^2 \psi_2+
\left(\Lambda-\alpha|\psi_1|^2 - |\psi_2|^2\right)\psi_2
\end{eqnarray}
\end{subequations}
together with the boundary conditions $| \psi_1|\rightarrow 1, \psi_2 \rightarrow 0,$ as $|{\bf x}|\rightarrow \infty.$ 
In the phase separation regime $\alpha=g_{12}/g_{11}=g_{12}/g_{22} > 1$. Here $\Lambda=\mu_2/\mu_1$ 
where $\mu_1$ and $\mu_2$ are the dimensional chemical potentials of $\psi_1$ and $\psi_2$.
We solve numerically the discretized version of Eqs. (\ref{ueq}) by a Newton-Raphson algorithm 
 combined with a secant algorithm to find
$\Lambda$ for a given constraint on  
$N_2=\int |\psi_2|^2\, dx\, dy$. We obtain a family of solutions characterized by the velocity
of propagation $U$, energy, $E$ and impulse ${\bf p}=(p_1+p_2,0)$, given by
\begin{subequations}
\begin{eqnarray}
E  & = &  \int \left[\frac{1}{2}|\nabla\psi_1|^2 +\frac{1}{2}|\nabla\psi_2|^2 +
\alpha|\psi_1|^2|\psi_2|^2 \right.\\
& & +\left. \frac{1}{2}(1-|\psi_1|^2)^2 +
\frac{1}{2}|\psi_2|^4-\Lambda|\psi_2|^2 \right] dxdy,\label{E} \\
 p_1  & = & \text{Im} \left[  \int_{\mathbb{R}^2}  \left(\psi_1^*-1\right)\frac{\partial \psi_1}{\partial x}  \right] dxdy,\\ 
p_2 & =  &  \text{Im} \left[   \int_{\mathbb{R}^2} \ \psi_2^*\frac{\partial \psi_{2}}{\partial x}  dxdy \right].
\end{eqnarray}
\end{subequations}
The resulting families of
solutions are given on Fig. \ref{fig_disp} for various choices of $N_2$ 
together with the Jones-Roberts (JR) dispersion relations \cite{jr} for one-component condensates. For a given speed $U$, solitary solutions with higher $\alpha$  have lower energy and higher impulse. 
\begin{figure}
\epsfig{file=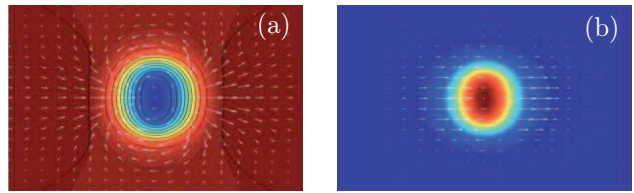,width=0.96\columnwidth}
\caption{(Color online) A stationary bound state solution for $U=0.2$, $N_2 = 80$. (a) $|\psi_1(x,y)|^2$ (b) $|\psi_2(x,y)|^2$. Shown are pseudocolor plots of the densities, arrows indicating the direction of the local current density $\boldsymbol{j} =- \text{Re}\left( i \psi \nabla \psi^*\right)$ and for (a) a few contour lines (black solid lines). The spatial region shown in the plot is  $(x,y) \in [-30,30]\times [-20,20]$}
\label{fig_density}
\end{figure}
Firstly,
in contrast with the JR solutions, there is a stationary solitary wave
corresponding to the ground state of the system with
all the mass of the second component forming a radially symmetric
``bubble'' in the centre of the depleted first component. This complex
has a nonzero energy $E_{gs}$.
As the
velocity increases from zero, the bubble becomes oblate in the direction of the motion with the velocity field of the 
first component is that of a dipole as Fig. \ref{fig_density} illustrates.  
There is a point on the dispersion curve where the velocity
reaches its maximum -- the inflection point. As energy and momentum
continue to increase, the velocity decreases and the solutions become pairs of
vortices of opposite circulation in the first component with the
second component condensing in the vortex cores. In general the bubble-like solutions, for small $E$ can be seen as a bound state of  a JR rarefaction pulse and a mobile ``filling" of the second component. Fig. \ref{fig_disp}(b) shows that there is a maximum velocity for the propagation of these solutions well separated from the sound speed velocity what can be physically interpreted as a signature of the mass of the second component, the heavier being the second component the smaller being this maximum velocity.
\begin{figure}
 \epsfig{file=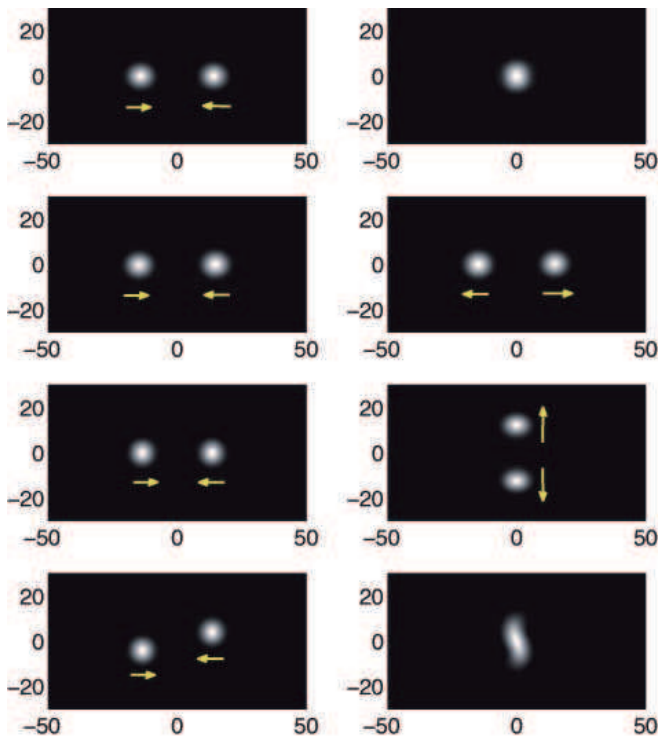,width=\columnwidth}
\caption{Outcomes of coherent bubble-droplet pairs collisions. Density snapshots of the second component are shown before (left column) and after (right column) the collision. In the first row, identical incident bubbles with $U=0.1, N_2=40, p=6.36,  E=2.99$ form a bound
 state. In the second row, for $U_1=U_2=0.2$ incident bubbles with the same speed but
 different sizes emulate a elastic collision (left bubble: $N_2=40, E=5.6, p=14.9$, right bubble: $N_2=50, E=7.7, p=20.1 $). In the third row,
 identical bubbles (parameters being $U=0.2, N_2=40, E=5.6, p=14.9$) emulate vortex
 recombination. In the fourth row, identical bubbles (same as those in
 third row) collide with an offset of 8 dimensionless units and form a bound state.}
\label{fig_collision}
\end{figure}
To study the collisional properties of these localized structures we
have simulated the evolution of two bubbles set on a colliding
course. Initially the bubbles are separated by a large distance, so
that individually they are accurately represented by the solutions we
found. We have observed several possible outcomes of such collisions summarized in Fig. \ref{fig_collision}. Almost identical slow colliding bubbles may form a bound stationary state  even when they collide with an offset. Bubbles moving with large velocities may scatter at $\pi/2$ angle resembling the collision of two pairs of vortices of opposite circulation. Almost elastic collisions between these
structures were observed  when the velocities or masses of the bubbles
were very different. We have observed that a bound state is more likely to  be formed when bubbles have similar phases of the second component and move slowly. 
 In all such collisions, a small fraction of the total mass is lost  and carried away by sound waves. The outgoing bubbles
are solitary wave solutions as verified by energy-impulse calculations.

\emph{Conclusions.-} We have discussed the dynamics of interacting bubble-droplet pairs in
immiscible Bose-Einstein condensates. Coherent atomic
soliton molecules made up of dark-bright soliton pairs with appropriate phases can be constructed in 1D scenarios. In 
two-dimensions, we have found novel types of robust bubble-like solitons without a scalar counterpart 
that can be seen as a bound state of two vector objects. Our ideas can be used to construct
 coherent atomic molecules made up of solitons.

\acknowledgments

\emph{Acknowledgements.-} This work has been supported by grants
FIS2007-29090-E (Ministerio de 
Ciencia e Innovaci\'on, Spain), PGIDIT04TIC383001PR (Xunta de Galicia) and PCI-08-0093
(Junta de Comunidades de Castilla-La Mancha, Spain).  NGB acknowledges support from EPSRC-UK and Isaac Newton Trust.
VAB acknowledges support from the FCT grant, PTDC/FIS/64647/2006.
We want to acknowledge David Novoa (U. Vigo) for discussions.

\end{document}